\documentclass[a4paper,DIV14, 12pt] {scrartcl}

\usepackage{fancyhdr}
\usepackage{isolatin1}
\usepackage{ltxtable}
\usepackage{textcomp}
\usepackage{tabularx}
\usepackage{graphicx, color}
\usepackage{subfigure}
\usepackage{amsmath}
\usepackage{amsfonts}
\usepackage{amssymb}
\usepackage{amsthm}
\usepackage{empheq}
\usepackage{fixmath}
\usepackage{tensind}
\usepackage{wasysym}  
\usepackage{afterpage}
\usepackage{dsfont}   
\usepackage{bibgerm}
\usepackage{hyperref}

\def\e{\mathrm{e}}
\def\ii{\mathrm{i}}

\def\bea{\begin{eqnarray}}
\def\eea{\end{eqnarray}}
\def\qbin#1#2{\genfrac{[}{]}{0pt}{}{#1}{#2}}

\def\psg{\phantom{-}}

\whenindex{w}{\mathbf{a}}{}
\whenindex{x}{\mathbf{b}}{}
\whenindex{y}{\mathbf{c}}{}
\whenindex{z}{\mathbf{d}}{}
\whenindex{W}{\mathbf{A}}{}
\whenindex{X}{\mathbf{B}}{}
\whenindex{Y}{\mathbf{C}}{}
\whenindex{Z}{\mathbf{D}}{}
\whenindex{'}{'}{}

\newcommand{\ket}[1]{|#1\rangle}

\def\bsm{\left( \!\begin{smallmatrix}}
\def\esm{\end{smallmatrix} \!\right)}

\DeclareMathOperator{\rog}{L}

\tensordelimiter{?}

\numberwithin{equation}{section}

\newcommand{\Z}{\mathbb{Z}}

\newcommand{\N}{\mathbb{N}}

\newcommand{\Tr}{\mbox{Tr}}
 
\newcommand{\2}{\frac{1}{2}}

\newcommand{\be}{\begin{equation}}
\newcommand{\ee}{\end{equation}}

\renewcommand{\pmod}[1]{\ (\mathrm{mod} \ #1)}
\newcommand{\vin}[1]{\in (\Z_{\geq 0})^{#1}}

\begin{document}
\selectlanguage{english}
\begin{titlepage}
\setcounter{page}{0}
\begin{flushright}
      hep-th/0611241\\
      ITP--UH-25/06\\
\end{flushright}
\vskip 2.0cm

\begin{center}
{\LARGE\bf Fermionic Expressions for the Characters of $\mathbold{c_{p,1}}$ Logarithmic Conformal Field Theories}
\\
\vspace{14mm}

{\Large
Michael Flohr}
\ \ \ and \ \ \ 
{\Large
Carsten Grabow}
\ \ \ and \ \ \ 
{\Large
Michael Koehn}
\\[5mm]
{ \em
Institut f\"ur Theoretische Physik, Leibniz-Universit\"at Hannover \\
Appelstra\ss{}e 2, 30167 Hannover, Germany }
\\[5mm]
email: \texttt{flohr / grabow / koehn @ itp.uni-hannover.de}
\\[12mm]
\small \today
\end{center}
\vspace{15mm}

\begin{abstract}
We present fermionic quasi-particle sum representations consisting of a single fundamental fermionic form for all characters of the logarithmic conformal field theory models with central charge $c_{p,1}$, $p \geq 2$, and suggest a physical interpretation. We also show that it is possible to correctly extract dilogarithm identities.\\
\phantom{-}\\ \phantom{-}
\end{abstract}
\vfill
\end{titlepage}

\pagenumbering{arabic} \setcounter{page}{1}

\section{Introduction}

Infinite dimensional symmetry algebras play a major role in two-dimensional conformal field theory (CFT). In particular, the Hilbert spaces of the best known CFTs, the minimal models \cite{BPZ84},
decompose into irreducible highest weight representations of such algebras. More generally, the characters of the representations are an essential ingredient for a CFT. They can be written as $\chi(q)=q^{-\frac{c}{24}}\Tr q^{L_0}$, with $c$ being the central charge and $L_0$ the Virasoro zero mode, and hence encode the energy spectrum (at least certain sectors). The trace is usually taken over an irreducible highest weight representation and the factor $q^{-\frac{c}{24}}$ guarantees the needed linear behavior under modular transformations. 

The minimal models are distinguished by the central charge, which can be parameterized by two coprime integers $p$ and $p'$ (w.l.o.g. we set $ p>p' $) as
\be
c_{p,p'}=1-6\frac{(p-p')^2}{pp'} \ ,
\ee
and the highest weights 
\be 
\label{highestweight}
h^{p,p'}_{r,s}=\frac{(pr-p's)^2-(p-p')^2}{4pp'} 
\ee
with $ 1 \le r < p' $ and $ 1 \le s < p $. Their characters are constructed as formal power series $\chi$ in some variable $ q=\e^{2\pi \ii \tau}, \tau\in\mathfrak{h} $ (upper half-plane).
Non-unique bases of the Hilbert spaces in two dimensional CFTs establish the existence of several alternative character formulae. From both a mathematical and physical point of view, further interest attaches to the so-called \emph{fermionic sum representations} for a character, which first appeared in the context of the Rogers-Schur-Ramanujan identities \cite{Rog94,Sch17,RR19} (for $ a\in\{0,1\} $) 
\be
\label{rsr}
\sum\limits_{n=0}^{\infty}\frac{q^{n(n+a)}}{(q)_n} = \prod\limits^{\infty}_{n=1}
\frac{1}{(1-q^{5n-1-a})(1-q^{5n-4+a})}
\ee
with
\be
(q)_n=\prod\limits_{i=1}^{n}
(1-q^i) \quad \text{and per definition} \quad  (q)_0=1\quad \text{and} \quad (q)_{\infty}=\lim_{n\rightarrow\infty}(q)_n \ .
\ee
These identities coincide with the two characters of the $ \mathcal{M} (5,2) $ minimal model (normalized to 1 at $q=0$). By using Jacobi's triple product identity (see e.g. \cite{And84}), the r.h.s. of \eqref{rsr} can be transformed to give a simple example of what is called a \emph{bosonic-fermionic $q$-series identity}:
\be \label{bf_id}
\sum\limits_{n=0}^{\infty}\frac{q^{n(n+a)}}{(q)_n} =  \frac{1}{(q)_{\infty}}\sum\limits_{n=-\infty}^{\infty}(q^{n(10n+1+2a)}-q^{(5n+2-a)(2n+1)}) \ .
\ee
The so-called bosonic expression on the r.h.s. of \eqref{bf_id} corresponds to two special cases of the general character formula for minimal models $ \mathcal {M}  (p,p') $ \cite{RoC84} 
\be
\label{generalcharacter}
\hat{\chi}^{p,p'}_{r,s}=q^{\frac{c}{24}-h^{p,p'}_{r,s}}\chi^{p,p'}_{r,s}=\frac{1}{(q)_{\infty}}\sum\limits_{n=-\infty}^{\infty}(q^{n(npp'+pr-p's)}-q^{(np+s)(np'+r)})
\ee 
with $ \hat{\chi}^{p,p'}_{r,s} $ being the normalized character. The symmetry $ \chi^{p,p'}_{r,s}=\chi^{p,p'}_{p'-r,p-s} $ follows from \eqref{highestweight}. Since \eqref{generalcharacter} is computed by eliminating null states from the Hilbert space of a free chiral boson \cite{FeF83}, it is referred to as bosonic form. Its signature is the alternating sign, which reflects the subtraction of null vectors. The factor $ (q)_\infty $ keeps track of the free action of the Virasoro 'raising' modes. Furthermore, it can be expressed in terms of $ \Theta $-functions \eqref{theta}, which directly point out the modular transformation properties of the character.

In contrast, the fermionic sum representation possesses a remarkable interpretation in terms of quasi-particles for the states, obeying Pauli's exclusion principle.
In the first systematic study of fermionic expressions \cite{KKMM93b}, sum representations for all characters of the unitary Virasoro minimal models and certain non-unitary minimal models were given.
The list of expressions was augmented to all $p$ and $p'$ and certain $r$ and $s$ in \cite{BMS98}.  Eventually, the fermionic expressions for the characters of all minimal models were presented in \cite{Wel05}. 
Such a fermionic expression, which is a generalization of the left hand side of \eqref{rsr}, is a linear combination of fundamental fermionic forms.
A \emph{fundamental fermionic form} \cite{BMS98,Wel05,DKMM94} is 
\be
\label{fff}\sum_{\substack{\vec{m}\in(\mathbb{Z}_{\geq 0})^r \\ \text{restrictions}}}\frac{q^{\vec{m}^tA\vec{m}+\vec{b}^t\vec{m}+c}}{\prod_{i=1}^{j}(q)_i}\prod_{i=j+1}^{r}\qbin{g(\vec{m})}{m_i}_q
\ee
with $A\in M_r(\mathbb{Q})$, $\vec{b}\in\mathbb{Q}^r$, $c\in\mathbb{Q}$, $0 \leq j \leq r$, $g$ a certain linear, algebraic function in the $m_i, 1\leq i \leq r$, and the $q$-binomial coefficient defined as 
\be
\begin{bmatrix} n \\ m \end{bmatrix}_q=\begin{cases} \frac{(q)_{n}}{(q)_{m}(q)_{n-m}} & \mbox{if} \; \; 0 \le m \le n \; \; \\ 0 & \mbox{otherwise} \; \; \\ \end{cases} .
\ee
If $j=r$, then the fundamental fermionic form reduces to the form that is found in \emph{Nahm's conjecture} (see e.g. \cite{Nah04})
\be \label{nahm}
f_{A,\vec{b},c}(\tau)=\sum_{\substack{\vec{m}\in(\Z_{\geq 0})^r \\ \text{restrictions}}}\frac{q^{\vec{m}^tA\vec{m}+\vec{b}^t\vec{m}+c}}{(q)_{\vec{m}}} \ ,
\ee
which makes a prediction whether for a given matrix $A$ there exist $\vec{b}$ and $c$ such that \eqref{nahm} is a modular function.\footnote{The constant $c$ is not to be confused with the central charge $ c_{p,p'} $.}
The bosonic representations are in general unique, whereas there is usually more than one fermionic expression for the same character.

In addition to the noted minimal models, there exist other theories that have more symmetries than just the Virasoro algebra. They are generated by modes of currents different from the energy-momentum tensor. Possible extensions lead to free fermions, Ka\v{c}-Moody algebras, Superconformal algebras or $\mathcal{W}$-algebras. Here we focus on characters of representations of these extended symmetry algebras (especially of the $\mathcal{W}$-algebras), which contain the Virasoro algebra as a subalgebra. Specifically, we are interested in the $\mathcal{W}(2,2p-1,2p-1,2p-1)$ series of so-called triplet algebras \cite{Kau91}, which constitute the best understood examples of logarithmic conformal field theory (LCFT) models\footnote{For reviews, see \cite{Flo03,Gab03} and references therein.} and have central charges $c_{p,1}=1-6\frac{(p-1)^2}{p}$. For LCFTs, almost all of the basic notions and tools of (rational) CFTs, such as null vectors, (bosonic) character functions, partition functions, fusion rules, modular invariance, have been generalized by now. The main difference to ordinary rational
CFTs such as the minimal models is the occurence of indecomposable
representations. By contributing a complete set of fermionic character expressions for the $\mathcal{W}(2,2p-1,2p-1,2p-1)$ models with $p\geq 2$ (which we refer to as the $c_{p,1}$ models), we provide further evidence to answer the question about whether these models, although they lie outside the usual classification scheme of rational CFTs, are nonetheless bona fide theories.

Furthermore, as mentioned above, fermionic sum representations for characters admit an interpretation in terms of fermionic quasi-particles, as shown in \cite{KM93} (see also \cite{KKMM93a}). This can be easily seen from the sum \eqref{nahm} with the help of combinatorics: The number of additive partitions $ P_M(N,N') $ of a positive integer $ N $ into $ M $ distinct\footnote{The requirement of distinctiveness expresses the fermionic nature of the quasi-particles, i.e. Pauli's exclusion principle.} non-negative integers which are smaller than or equal to $ N' $ is stated by \cite{Sta72}
\be
\sum\limits_{N=0}^{\infty}P_M(N,N')q^N=q^{\frac{1}{2}M(M-1)}\begin{bmatrix}N'+1 \\ M \end{bmatrix}_q \ ,
\ee
which in the limit $ N' \to \infty $ takes the form
\be \label{distinct_intro}
\lim_{N'\to\infty}\sum\limits_{N=0}^{\infty}P_M(N,N')q^N=q^{\frac{1}{2}M(M-1)}\frac{1}{(q)_M} \ .
\ee
Applying \eqref{distinct_intro} to the fermionic sum representation \eqref{nahm} leads to 
\be \label{zwischen_intro}
\prod_{i=1}^r \left(\sum_{\substack{m_i\\ \text{restrictions}}}^{\infty}\sum_{N=0}^{\infty}P_{m_i}(N)q^{N+(b_i+\frac{1}{2})m_i+\sum_{j=1}^r A_{ij}m_im_j-\frac{1}{2}m_i^2}\right) \ ,
\ee
where $P_M(N)=\lim_{N'\to\infty}P_M(N,N')$. The constant $c$ has been omitted, since it would just result in an overall shift of the energy spectrum. For the quasi-particle interpretation, the characters are regarded as partition functions $ Z $ for left-moving excitations with the ground-state energy scaled out
\be \label{partition_function}
\chi \sim Z=\sum\limits_{\text{states}}\e^{-\frac{E_{\text{states}}}{kT}}=\sum\limits_{l=0}^{\infty}P(E_l)\e^{-\frac{E_l}{kT}}
\ee
with $ T $ being the temperature, $ k $ the Boltzmann's constant, $ E_l $ the energy and $ P(E_l) $ the degeneracy of the particular energy level $ l $.
This means that if we assume massless single-particle energies
\be \label{single-particle-energy}
e(p^i_{\alpha})=vp^i_{\alpha}
\ee
($ v $ referred to as the fermi velocity, spin-wave velocity, speed of sound or speed of light), where $p_{\alpha}^i$ denotes the quasi-particle $\alpha$ of 'species' $i\ (1\leq i \leq r)$, and if in \eqref{zwischen_intro} we set 
\be \label{boltzmann}
q=\e^{-\frac{v}{kT}} \ ,
\ee
we deduce that the partition function corresponds to a system of quasi-particles that are of $r$ different species and which obey the Pauli exclusion principle
\be
p_{\alpha}^i \neq p_{\beta}^i  \quad \text{for} \ \alpha \neq \beta \quad \text{and all}\quad i \ ,
\ee
but whose momenta $p_{\alpha}^i$ are otherwise freely chosen from the sets
\be
P_i=\left\lbrace  p_{\text{min}}^i, p_{\text{min}}^i+1,p_{\text{min}}^i+2,\ldots,p_{\text{max}}^i \right\rbrace
\ee
with minimum momenta
\be
\label{pmin}p_{\text{min}}^i(\vec m)=\left[((A-\frac{1}{2})\vec m)_{i}+b_{i}+\frac{1}{2}\right]
\ee
and with the maximum momenta $p_{\text{max}}^i$ either infinite if $i \leq j$ in \eqref{fff} or, if $i>j$, finite and dependent on $\vec{m}$, $\vec{b}$ and $g$. Since the fermionic character expressions we present for the $c_{p,1}$ series of LCFTs are all of the type \eqref{nahm}, we will only deal with the case that all $p_{\text{max}}^i=\infty$ in this article, i.e. the spectra are not bounded from above.
This means that a multi-particle state with energy $E_l$ may consist of exactly those combinations of quasi-particles of arbitrary species $i$, whose single-particle energies $e(p^i)$ add up to $E_l$ and where Pauli's principle holds for any two quasi-particles of that combination unless they belong to different species. Possible sum restrictions then result in the requirement that certain particles may only be created in conjunction with certain others.

The outline of the article is as follows: In section 2, the bosonic character expressions for the LCFT models corresponding to central charge $c_{p,1}$ are given and the fermionic counterparts are obtained. In section 3, it is shown that the new fermionic character expressions correctly lead to dilogarithm identities. In section 4, we provide a physical interpretation of the results of section 2.
\mathversion{bold}
\section{Characters of the Triplet Algebras $\mathcal{W}(2,2p-1,2p-1,2p-1)$}
\mathversion{normal}

\subsection{Highest Weight Representations} 
A minimal model with central charge $ c_{p,p'} $ admits highest weights $h_{r,s}^{p,p'}$ (see \eqref{highestweight}) with $ 1 \le r < p' $ and $ 1 \le s < p $. In contrast, the $h$-values of all $3p-1$ inequivalent representations for the LCFT models with $c_{p,1}$ and chiral symmetry algebra $\mathcal{W}(2,2p-1,2p-1,2p-1)$ can be read off the extended conformal grid of the augmented minimal model \cite{Flo97,EF06b}, corresponding formally to central charge $c_{3p,3}$.
For example, in the case $p=2$ and $c_{2,1}=-2$, the only possible highest weights are $h \in \{ -\frac{1}{8},0,\frac{3}{8},1 \}$, where $h=0$ corresponds to two inequivalent representations \cite{GK96b}.
In comparison to the singlet algebra $\mathcal{W}(2,2p-1)$, which is too small to obtain a rational $ c_{p,1} $ model, the triplet algebra now serves as its maximally extended symmetry algebra.
The way to get the $\mathcal{W}$-algebra characters is to sum up appropriate subsets of Virasoro characters of degenerate highest weight representations, keeping in mind that only those highest weights are permitted which differ by integers and taking care of multiplicities caused by the $\mathfrak{su}(2)$ symmetry among the triplet of chiral fields of conformal
weight $ 2p-1 $.

\subsection{Characters in Bosonic Form}

Analyzing the action of the triplet algebras on the degenerate Virasoro representations \cite{GK96b,Flo96} as well as the modular transformation properties of the vacuum character allows to find a complete set of character functions for the $c_{p,1}$ models that is closed under modular transformations \cite{Flo97}:
\begin{align}
\label{triplet-char-0}\chi_{0,p} & = \frac{\Theta_{0,p}}{\eta} & \text{\normalsize \textrm{representation to} } \qquad & {\textstyle h_{1,p}^{p,1}} \\
\chi_{p,p} & = \frac{\Theta_{p,p}}{\eta} & & {\textstyle h_{1,2p}^{p,1}} \\
\chi_{\lambda,p}^{+} & = \frac{(p-\lambda)\Theta_{\lambda,p}+(\partial\Theta)_{\lambda,p}}{p\eta} & & {\textstyle h_{1,p-\lambda}^{p,1}} \\
\chi_{\lambda,p}^{-} & = \frac{\lambda\Theta_{\lambda,p}-(\partial\Theta)_{\lambda,p}}{p\eta} & & {\textstyle h_{1,3p-\lambda}^{p,1}} \\
\label{tilde_char_plus}\tilde{\chi}_{\lambda,p}^{+} & = \frac{\Theta_{\lambda,p}+\ii\alpha\lambda(\nabla\Theta)_{\lambda,p}}{\eta} & & {\textstyle h_{1,p+\lambda}^{p,1}} \\
\label{tilde_char_minus}\tilde{\chi}_{\lambda,p}^{-} & = \frac{\Theta_{\lambda,p}-\ii\alpha(p-\lambda)(\nabla\Theta)_{\lambda,p}}{\eta} & & {\textstyle h_{1,p+\lambda}^{p,1}}
\end{align}
where $0<\lambda<p$, $k=pp'=p$, $\lambda=pr-p's=pr-s$ and with the \emph{Jacobi-Riemann $\Theta$-function} defined as
\be 
\label{theta}\Theta_{\lambda,k}(\tau)=\sum\limits_{n\in \mathbb Z}q^{\frac{(2kn+\lambda)^2}{4k}} \ ,
\ee
the \emph{affine $\Theta$-function} defined as
\be
\label{deltheta}(\partial \Theta)_{\lambda,k}(\tau)=\sum\limits_{n\in \mathbb Z}(2kn+\lambda)q^{\frac{(2kn+\lambda)^2}{4k}}
\ee
and the Dedekind $ \eta $-function defined as 
\be
\eta(q)=q^{1/24}(q)_\infty \ .
\ee
Here, $q=\e^{2\pi \ii \tau}, \tau\in\mathfrak{h}$ (upper half-plane), $\lambda$ is called the \emph{index} and $k$ the \emph{modulus}. The $\Theta$-functions satisfy the symmetries 
\begin{gather}
\label{theta_symmetries}\Theta_{\lambda,k}=\Theta_{-\lambda,k}=\Theta_{\lambda+2k,k} \\
\label{deltheta_symmetries}(\partial\Theta)_{-\lambda,k}=-(\partial\Theta)_{\lambda,k} \ .
\end{gather}
$\frac{\Theta_{\lambda,k}(\tau)}{\eta(\tau)}$ is a modular form of weight zero with respect to the generators $\mathcal{T}:\ \tau \mapsto \tau +1$ and $\mathcal{S}:\ \tau \mapsto -\frac{1}{\tau}$ of the modular group $PSL(2,\mathbb{Z})$. But since $\frac{(\partial\Theta)_{\lambda,k}(\tau)}{\eta(\tau)}$ is a modular form of weight one with respect to $\mathcal{S}$, some of the above character functions are of inhomogeneous modular weight, thus leading to $S$-matrices with $\tau$-dependent coefficients. However, adding
\be
(\nabla \Theta )_{\lambda,k}(\tau) = \frac{\log q}{2 \pi \ii}\sum_{n\in \mathbb Z}(2kn+\lambda)q^{\frac{(2kn+\lambda)^2}{4k}}\ ,
\ee
one finds a closed finite dimensional representation of the modular group with constant $S$-matrix coefficients.

Note that \eqref{tilde_char_plus} and \eqref{tilde_char_minus} are not characters of representations in the usual sense. Actually, these are regularized character functions and the $\alpha$-dependent part has an interpretation as torus vacuum amplitudes \cite{FG06}. In the limit $\alpha\rightarrow 0$, they become the characters of the full reducible but indecomposable representations.
\mathversion{bold}
\subsection{Fermionic Character Expressions for $\mathcal{W}(2,3,3,3)$}
\mathversion{normal}
We present fermionic sum representations for all characters of all $c_{p,1}$ models. All of them consist of only one fundamental fermionic form. In this section, we present in detail the fermionic expressions for the case $p=2$ and then generalize to $p>2$ in the next section.

In the case of $p=2$, the bosonic characters read:
\begin{align}
\label{h11}\chi_{1,2}^{+} & = \frac{\Theta_{1,2}+(\partial\Theta)_{1,2}}{2\eta} 
& {\textstyle \text{\normalsize \textrm{vacuum irrep }} V_0\text{ \normalsize \textrm{to} } h_{1,1}=0} \\
\label{h12}\chi_{0,2} & = \frac{\Theta_{0,2}}{\eta} & {\textstyle \text{\normalsize \textrm{irrep to} } h_{1,2}=-\frac{1}{8}}\\
\label{tilde_char_simplified}\chi_{1,2} & = \frac{\Theta_{1,2}}{\eta} 
& {\textstyle \text{\textrm{\normalsize indecomp. rep }} R_0 ( \supset V_0 ) \text{ \normalsize \textrm{to} } h_{1,3}=0 } \\
\label{h14}\chi_{2,2} & = \frac{\Theta_{2,2}}{\eta} & {\textstyle \text{\textrm{\normalsize irrep to} } h_{1,4}=\frac{3}{8}}\\
\label{h15}\chi_{1,2}^{-} & = \frac{\Theta_{1,2}-(\partial\Theta)_{1,2}}{2\eta} & {\textstyle \text{\textrm{\normalsize irrep to} } h_{1,5}=1} & .
\end{align}
When $\alpha \rightarrow 0$, the general forms \eqref{tilde_char_plus} and \eqref{tilde_char_minus} lead to the character expression \eqref{tilde_char_simplified} \cite{Kau95,Flo97}. Actually, there exist two indecomposable representations, $R_0$ and $R_1$, which, however, are equivalent and thus share the same character.

The fermionic expression for $\frac{\Theta_{\lambda,k}(\tau)}{\eta(\tau)}$ can be calculated from its bosonic counterpart via the Durfee rectangle identity
\be
\sum_{n=0}^{\infty}\frac{q^{n^2+nk}}{(q)_n(q)_{n+k}}=\frac{1}{(q)_{\infty}}
\ee 
(see e.g. \cite{And84}) to be the sum-restricted $r$-fold \sloppy \mbox{$q$-hypergeometric} series
\begin{align}
\label{fe-theta_by_eta}\Lambda_{\lambda,k}(\tau) &  =  \frac{\Theta_{\lambda,k}(\tau)}{\eta(\tau)} \nonumber \\
 & = \sum_{\substack{\vec{m}\vin{2} \\ m_1+m_2\equiv 0  \pmod{2}}}\frac{q^{\frac{1}{4}\vec{m}^t\bsm k & 2-k\\2-k & k \esm\vec{m}+\frac{1}{2}\bsm \psg \lambda\\-\lambda \esm^{\!{\!t}} \vec{m}+\frac{\lambda^2}{4k}-\frac{1}{24}}}{(q)_{\vec{m}}}  \nonumber \\
 & = \sum_{\substack{\vec{m}\vin{2} \\ m_1+m_2\equiv 1 \pmod{2}}}\frac{q^{\frac{1}{4}\vec{m}^t\bsm k & 2-k\\2-k & k \esm \vec{m}+\frac{1}{2}\bsm -(k-\lambda)\\ \psg k-\lambda \esm^{\!{\!t}} \vec{m}+\frac{(k-\lambda)^2}{4k}-\frac{1}{24}}}{(q)_{\vec{m}}}
\end{align}
with $(q)_{\vec{m}}=\prod_{i=1}^{r}(q)_{m_i},\  r=2$ \cite{KMM93}.\footnote{Note that \eqref{fe-theta_by_eta} is not unique just as \eqref{theta}: According to \eqref{theta_symmetries}, the vector may be changed in certain ways along with the constant.} This serves for \eqref{h12} to \eqref{h14} and is in agreement with Nahm's conjecture (see e.g. \cite{Nah04}), which predicts that for a matrix of the form $A=\scriptstyle\frac{1}{2}\bsm\alpha & 1-\alpha \\ 1-\alpha & \alpha\esm$ with rational coefficients, there exist a vector $\vec{b}\in\mathbb{Q}^r$ and a constant $c\in\mathbb{Q}$ such that \eqref{nahm} is a modular function.

The fermionic expressions of the remaining two characters may be calculated as follows:
By using $\frac{(\partial\Theta)_{1,2}}{\eta^3(q)}=1$ and the easily proven identity
\be
\eta(q)=q^{\frac{1}{24}}\sum_{n=0}^{\infty}\frac{(-1)^nq^{{n+1\choose 2}}}{(q)_n}
\ee
by Euler  (see e.g. \cite{And84}), which implies that
\be
\label{fe-eta_squared}\eta^2(q)=\tilde{\eta}^2(q,-1) \quad \text{with} \quad \tilde{\eta}^2(q,z)=\sum_{\vec{m}\vin{2}}\frac{q^{\frac{1}{2}\vec{m}^t\bsm 1 & 0\\0 & 1 \esm \vec{m}+\frac{1}{2}\bsm 1\\1 \esm^{\!{\!t}} \vec{m}+\frac{1}{12}}z^{m_1+m_2}}{(q)_{\vec{m}}} \ ,
\ee
and by using furthermore the relation
\be
\sum_{\vec{m}\vin{2}}\frac{q^{\frac{1}{2}\vec{m}^t\bsm 1 & 0\\0 & 1\esm\vec{m}+\frac{1}{2}\bsm 1\\1\esm^{\!{\!t}} \vec{m}}}{(q)_{\vec{m}}}=\sum_{\substack{\vec{m}\vin{2} \\ m_1+m_2\equiv 0 \pmod{2}}}\frac{q^{\frac{1}{2}\vec{m}^t\bsm 1 & 0\\0 & 1 \esm\vec{m}+\frac{1}{2}\bsm \psg 1\\-1 \esm^{\!{\!t}} \vec{m}}}{(q)_{\vec{m}}}
\ee
(which follows essentially from $\sum_{m=0}^{\infty}\frac{q^{m(m+1)/2}}{(q)_m}=\frac{1}{2}\sum_{m=0}^{\infty}\frac{q^{m(m-1)/2}}{(q)_m}$), the remaining two characters also yield expressions which consist of only one fundamental fermionic form. 

The following is a list of the fermionic expressions for all five characters of the LCFT model corresponding to central charge $c_{2,1}=-2$:
\begin{align}
\label{fe-h11}\chi_{1,2}^{+} & = \sum_{\substack{\vec{m}\vin{2} \\ m_1+m_2\equiv 0 \pmod{2}}}\frac{q^{\frac{1}{2}\vec{m}^t\bsm 1 & 0\\0 & 1 \esm\vec{m}+\frac{1}{2}\bsm 1\\ 1 \esm^{\!{\!t}}\vec{m}+\frac{1}{12}}}{(q)_{\vec{m}}} \\
\label{fe-h12}\chi_{0,2} & = \sum_{\substack{\vec{m}\vin{2} \\ m_1+m_2  \equiv 0 \pmod{2}}}\frac{q^{\frac{1}{2}\vec{m}^t\bsm 1 & 0\\0 & 1 \esm\vec{m}-\frac{1}{24}}}{(q)_{\vec{m}}} \\
\label{fe-h13}\chi_{1,2} & = \sum_{\substack{\vec{m}\vin{2} \\ m_1+m_2\equiv 0 \pmod{2}}}\frac{q^{\frac{1}{2}\vec{m}^t\bsm 1 & 0\\0 & 1 \esm\vec{m}+\frac{1}{2}\bsm \psg 1 \\-1 \esm^{\!{\!t}}\vec{m}+\frac{1}{12}}}{(q)_{\vec{m}}} \\
\label{fe-h14}\chi_{2,2} & = \sum_{\substack{\vec{m}\vin{2} \\ m_1+m_2\equiv 0 \pmod{2}}}\frac{q^{\frac{1}{2}\vec{m}^t\bsm 1 & 0\\0 & 1 \esm\vec{m}+\bsm \psg 1\\-1\esm^{\!{\!t}}\vec{m}+\frac{11}{24}}}{(q)_{\vec{m}}} \\
\label{fe-h15}\chi_{1,2}^{-} & = \sum_{\substack{\vec{m}\vin{2} \\ m_1+m_2\equiv 1 \pmod{2}}}\frac{q^{\frac{1}{2}\vec{m}^t\bsm 1 & 0\\0 & 1 \esm\vec{m}+\frac{1}{2}\bsm 1 \\1 \esm^{\!{\!t}}\vec{m}+\frac{1}{12}}}{(q)_{\vec{m}}} \quad .
\end{align}
Using the equality to the bosonic representation of the characters, these give \emph{bosonic-fermionic q-series identities} generalizing the left and right hand sides of \eqref{bf_id}.
In \eqref{fe-h12} to \eqref{fe-h14}, also the last line of \eqref{fe-theta_by_eta} may be used, where $m_1+m_2 \equiv 1 \pmod{2}$.

It is remarkable that, although two of the bosonic characters have inhomogeneous modular weight, there is a uniform fermionic representation for all five characters with the same matrix $A$ in every case. But on the other hand, this is a satisfying result, since this is also the case for all other models for which fermionic character expressions are known: Their different modules are only distinguished by the linear term in the exponent, not by the quadratic one. Note that the fact that the quadratic form is diagonal fits well with the description of the $c=-2$ model in terms of symplectic fermions \cite{Kau95,Kau00}, see section \ref{quasi}.
\mathversion{bold}
\subsection{Fermionic Character Expressions for $p>2$}\label{generalize}
\mathversion{normal}

We now generalize the results of the foregoing section to $p>2$ and present fermionic sum representations for all characters of the LCFT models corresponding to central charge $c_{p,1}$. All of them consist of a single fundamental fermionic form.

The matrix $A$ in the case of $p=2$ can be understood as the degenerate inverse Cartan matrix of the series of Lie algebras $D_p$. Then, generalizing to the case $p>2$, we checked numerically up to $k=5$ to high order and assume it for $k>5$ that the fermionic character expressions for the $c_{p,1}$ models\footnote{This means the characters and not the torus vacuum amplitudes \eqref{tilde_char_plus} and \eqref{tilde_char_minus}. Note that $\lim\limits_{\alpha\to 0}\tilde{\chi}_{\lambda,k}^+=\lim\limits_{\alpha\to 0}\tilde{\chi}_{\lambda,k}^-=\chi_{\lambda,k}$ for $0<\lambda<k$.} can be expressed as follows and indeed equal the bosonic ones (cf. \eqref{triplet-char-0}-\eqref{tilde_char_minus}), the latter being redisplayed on the r.h.s. for convenience:
\begin{alignat}{2}
\label{fe-1}\chi_{\lambda,k} & = \sum_{\substack{\vec{m}\vin{k} \\ m_{k-1}+m_k\equiv 0 \pmod{2}}}\mspace{-8mu} \frac{q^{\vec{m}^t  C_{D_{k}}^{-1}\vec{m}+\vec{b}_{\lambda,k}^t\vec{m}+c^{\star}_{\lambda,k}}}{(q)_{\vec{m}}} && = \frac{\Theta_{\lambda,k}}{\eta} \\
\label{fe-2}\chi_{\lambda',k}^{+} & =  \sum_{\substack{\vec{m}\vin{k} \\ m_{k-1}+m_k\equiv 0 \pmod{2}}}\mspace{-8mu}\frac{q^{\vec{m}^t C_{D_{k}}^{-1}\vec{m}+\vec{b'}_{\lambda',k}^{+^{t}}\vec{m}+c^{\star}_{\lambda',k}}}{(q)_{\vec{m}}} && =  \frac{(k-\lambda')\Theta_{\lambda',k}+(\partial\Theta)_{\lambda',k}}{k\eta} \\
\label{fe-3}\chi_{\lambda',k}^{-} & =  \sum_{\substack{\vec{m}\vin{k} \\ m_{k-1}+m_k\equiv 1 \pmod{2}}}\mspace{-8mu}\frac{q^{\vec{m}^t C_{D_{k}}^{-1}\vec{m}+\vec{b'}_{\lambda',k}^{-^t}\vec{m}+c^{\star}_{k-\lambda',k}}}{(q)_{\vec{m}}} && = \frac{\lambda'\Theta_{\lambda',k}-(\partial\Theta)_{\lambda',k}}{k\eta}
\end{alignat}
for $0 \leq \lambda \leq k$ and $0 < \lambda' < k$, where $k=p$ since $p'=1$ and $(\vec{b}_{\lambda,k})_i=\frac{\lambda}{2}(\pm\delta_{i,k-1}\mp\delta_{i,k})$ for $1 \leq i \leq k$, $(\vec{b'}_{\lambda',k}^{+})_i=\max\{0,\lambda'-(k-i-1)\}$ for $1 \leq i < k-1$ and $(\vec{b'}_{\lambda',k}^{+})_i=\frac{\lambda'}{2}$ for $k-1 \leq i \leq k$, $(\vec{b'}_{\lambda',k}^{-})_i=(\vec{b'}_{k-\lambda',k}^{+})_i$ and $c^{\star}_{\lambda,k}=\frac{\lambda^2}{4k}-\frac{1}{24}$.\footnote{Note that in \eqref{fe-1}, also $m_{k-1}+m_k\equiv 1 \pmod{2}$ may be used as restriction, but then the vector and the constant change to $\vec{b}_{k-\lambda,k}$ and $c^{\star}_{k-\lambda,k}$, respectively (cf. \eqref{fe-theta_by_eta}).}
Thus, as in the previous section, the $p\times p$ matrix $A=C_{D_p}^{-1}$ is the same for all characters corresponding to a fixed $p$, i.e. for a fixed model. This is in agreement with previous results on fermionic expressions, since it is known to also be the case for the characters of a given minimal model (see e.g. \cite{Wel05}). Note that in \cite{KKMM93a}, fermionic expressions for the characters of the free boson with central charge $c=1$ and compactification radius $r=\sqrt{\frac{p}{2}}$ \cite{Gin88} have been obtained. Those characters equal \eqref{fe-1}. Thus, some of the expressions in \eqref{fe-1} already appeared in \cite{KKMM93a}, but only for $\lambda=0$ and $\lambda=k$ and only for the special case $ \vec{b}=\vec{0} $.
\subsection{More Fermionic Forms}
Some of the summands of the bosonic character expressions of the triplet algebras $\mathcal{W}(2,2p-1,2p-1,2p-1)$ resemble the Ka\v{c}-Peterson characters of the affine Lie algebra $ A^{(1)}_1 $ \cite{KP84}. Fermionic expressions for those characters are known, but most of them are not of type \eqref{nahm} but instead involve $q$-binomial coefficients. We display in short the known fermionic expressions and present new fermionic expressions of type \eqref{nahm} below.

Note the fermionic expressions for the irreducible integrable representations of $A^{(1)}_1$ at level $k-2$ \cite{BLS95}
\be
\label{deltheta-fe}\frac{(\partial\Theta)_{\lambda,k}(\tau)}{\eta^3(\tau)}=\sum_{\substack{m_1, \ldots ,m_{k-1}=0 \\ (\vec{m}')_i\equiv (\vec{Q}(\lambda ))_i \pmod{2}}}^{\infty}\frac{q^{\vec{m}^t B_k\vec{m}+c^{\sharp}_{\lambda,k}}}{(q)_{m_1}(q)_{m_2}}\prod_{i=3}^{k-1}\begin{bmatrix}\lceil (\frac{1}{2}(2-C_{A_{k-2}})\vec{m}')_{i-1}\rceil \\ m_i\end{bmatrix}_q
\ee
for $0<\lambda<k$ with $\vec{m}'^t=( m_1+m_2 , m_3 , m_4 , \hdots , m_{k-1} )$ and
\be
4B_k=C_k+C_{A_{k-1}} \; , \quad (C_k)_{ij}= \begin{cases} -1 & \mbox{if} \; \; i+j \; \; \mbox{is even and} \; \; i+j\le4 \\ 2 & \mbox{if} \; \; i+j \; \; \mbox{is odd and} \; \; i+j\le4 \\ 0 & \mbox{if} \; \; i+j>4 \end{cases} \quad ,
\ee
where $C_{A_k}$ is the Cartan matrix of the Lie algebra $A_k \cong s \ell_{k+1}$ and $c^{\sharp}_{\lambda,k}=\frac{2\lambda^2+k-2k\lambda}{8k}$. Given any $x \in \mathbb{R}$, $\lceil x \rceil$ and $\lfloor x \rfloor$ mean the next integer greater than or equal to $x$ and the next integer less than or equal to $ x $, respectively. The following restrictions hold for the sum variables: $(\vec{m}')_i=(\vec{Q}(\lambda ))_i \pmod 2$ with $\vec{Q}(\lambda )=((\sum_{j=0}^{\lfloor \frac{\lambda}{2}-1\rfloor}\delta_{i,\lambda-(2j+1)})_i : i \in \{ 1,\ldots,k-2 \})\in(\mathbb{Z}_2)^{k-2}$, i.e. $\vec{Q}(\lambda )$ is either of the form $(1,0,1,0,\ldots,1,0,0,0,\ldots,0)$ if $\lambda$ is odd or of the form $(0,1,0,1,\ldots,1,0,0,0\ldots,0)$ if $\lambda$ is even.\footnote{The number and the placement of entries $1$ in the latter vector may be changed in certain ways, but then an inner product $\vec{b}^t\vec{m}$ with the $k-1$-component vector $\vec{b}^t=(\frac{1}{2},-\frac{1}{2},0,\ldots,0)$ has to be added to the quadratic form in the numerator of \eqref{deltheta-fe}.} 

For $\lambda=1$ and $\lambda=k-1$, there exists another expression. In both cases, it consists of a single fundamental fermionic form without sum restrictions and has $2(k-2)$ different sum indices.

For $\lambda=1$, it reads \cite{FS93a}
\be
\label{deltheta-fe-lambda-1}
\frac{(\partial\Theta)_{1,k}(\tau)}{\eta^3(\tau)}=\sum_{\vec{m}\vin{2(k-2)}}\frac{q^{\frac{1}{2}\vec{m}^t(C_{A_2}\otimes C_{T_{k-2}}^{-1})\vec{m}+c^{\flat}_{1,k}}}{(q)_{\vec{m}}} \ ,
\ee
where $C_{A_2}$ is as above, $C_{T_k}^{-1}$ is the inverse of the $k \times k$ Cartan matrix of the tadpole graph\footnote{The $C_{T_k}$ Cartan matrix differs from the $C_{A_k}$ Cartan matrix only by a 1 instead of a 2 in the lower right corner.} and the constant $c^{\flat}_{\lambda,k}=\frac{\lambda^2}{4k}-\frac{1}{8}$.

For $\lambda=k-1$, we present the following fermionic expression:
\be
\label{deltheta-fe-lambda-k-1}
\frac{(\partial\Theta)_{k-1,k}(\tau)}{\eta^3(\tau)}=\sum_{\vec{m}\vin{2(k-2)}}\frac{q^{\frac{1}{2}\vec{m}^t(C_{A_2}\otimes C_{T_{k-2}}^{-1})\vec{m}+(\vec{a}_2 \otimes \vec{b}_{k-2})^t\vec{m}+c^{\flat}_{k-1,k}}}{(q)_{\vec{m}}}
\ee
with $\vec{a}_2^t=(1 , -1)$ and $\vec{b}_k^t=(1,2,3, \hdots , k)$. It has been checked numerically up to $k=4$ and high order and is assumed to hold for higher values of $k$.
\section{Dilogarithm Identities}

To support the fermionic character expressions we derived in section \ref{generalize}, we show in this section that it is possible to correctly extract dilogarithm identities from them. The effective central charge of the given LCFT model should be expressible as a sum of dilogarithm functions evaluated at certain algebraic numbers, where these numbers are determined by the matrix $A$ in the quadratic form in the exponent of the fermionic character expression.

Dilogarithm identities for the central charges and conformal dimensions exist for at least large classes of rational CFTs.
It is conjectured \cite{NRT93} that all values of the effective central charges occurring in non-trivial rational CFTs can be expressed as one of those rational numbers that consist of a sum of an arbitrary number of dilogarithm functions evaluated at algebraic numbers from the interval $(0,1)$.
Thus, the study of dilogarithm identities arising from CFTs, e.g. the set of effective central charges that can be expressed with a fixed number $N$ in \eqref{general_dilog_id}, gives further insight into the classification of all rational CFTs.

Dilogarithm identities are relations of the form
\be
\label{general_dilog_id}\frac{1}{\rog(1)}\sum_{i=1}^{N}\rog(x_i)=d
\ee
with $x_i$ an algebraic, $d$ a rational number, $N$ being the size of the matrix $A$ in the fermionic form, and $\rog$ being the Rogers dilogarithm (see e.g. \cite{Lew58,Lew81b}), defined for $0<x<1$ by
\be
\rog(x)=\sum_{n=1}^{\infty}\frac{x^n}{n^2}+\frac{1}{2}\log (x)\log (1-x) \ .
\ee
The Rogers dilogarithm admits an analytic continuation on the complex plane as a multivalued analytical function of $x$. The dilogarithm and its generalization, the polylogarithm, appear in a lot of branches of mathematics and physics (see e.g. \cite{Kir95}).

The effective central charge is a quantity originating from the properties of the CFT characters with respect to modular transformations. It is the same for all $p$ of the LCFTs corresponding to central charge $c_{p,1}$ and it is given by
\be
c_{\text{eff}}^{p,1}=c_{p,1}-24h_{\text{min}}^{p,1}=1 \ .
\ee

The $x_i$ in \eqref{general_dilog_id} are obtained by using the common saddle point analysis of the fermionic character (see e.g. \cite{NRT93}), implying that the place of $d$ in \eqref{general_dilog_id} is taken by the effective central charge of the conformal field theory in question. This leads to a set of algebraic equations
\be \label{alg_eqs}
x_i=\prod_{j=1}^{N}(1-x_j)^{A_{ij}+A_{ji}}
\ee
that determine the $x_i$, with $A=C_{D_p}^{-1}$ in the case of $\mathcal{W}(2,2p-1,2p-1,2p-1)$.

Although those $c_{p,1}$ theories are non-minimal models on the edge of the conformal grid, it is still possible (numerically solving \eqref{alg_eqs}) to correctly extract the well-known infinite set of dilogarithm identities
\be
\label{dilog_id}\frac{1}{\rog(1)}\left( 2\rog(\frac{1}{p})+\sum_{j=2}^{p-1}\rog(\frac{1}{j^2})\right)=1 \quad \forall \ p \geq 2
\ee
(which can be found in \cite{Kir92} and references therein). This supports the fermionic sum representations presented in section \ref{generalize} for the characters of the $\mathcal{W}(2,2p-1,2p-1,2p-1)$ triplet algebras.
\section{Quasi-Particle Interpretation} \label{quasi}
Our remarks on the $ c_{p,1} $ series with its underlying triplet $\mathcal{W}$-algebra are based on the quasi-particle interpretation \cite{KM93,KKMM93a,KKMM93b} for the minimal models, where the underlying symmetry algebra is the Virasoro algebra. 

\mathversion{bold}
\subsection{The $c=-2$ Model}
\mathversion{normal}
We start with the case $ p=2 $, i.e. $ c_{2,1}=-2 $. In contrast to the characters for the
minimal models, these characters are the traces over the representation modules of the triplet $\mathcal{W}$-algebra, instead of the Virasoro
algebra only. However, although highest weight states are labeled
by two highest weights in this case, $h$ and $w$ as the eigenvalues of
$L_0$ and $W_0$ respectively, we consider only the traces of the
operator $q^{L_0-\frac{c}{24}}$. It turns out that these $\mathcal{W}$-characters
are given as infinite sums of Virasoro characters, for example \cite{Flo96}
\be
\chi_{\ket{0}} = \sum_{k=0}^\infty (2k+1)\chi_{\ket{h_{2k+1,1}}}^{\text{Vir}} \ .
\ee

Let us now come to the vacuum character \eqref{fe-h11} for the $c_{2,1}$ model, which features the interesting sum restriction $ m_1+m_2\equiv 0 \pmod{2} $ expressing the fact that particles of type $ 1 $ and $ 2 $ must be created in pairs. Thus, the existence of one-particle states for either particle species is prohibited. Therefore, the single-particle energies must be extracted out of the observed multi-particle energy levels.

Applying \eqref{distinct_intro} to the fermionic sum representation \eqref{fe-h11} of the vacuum character leads to 
\be \label{zwischen}
\chi_{1,2}^+=\biggl( \sum_{m_1=0}^{\infty}\sum_{N=0}^{\infty}P_{m_1}(N)q^{N+m_1}\biggr) \biggl( \sum_{\substack{m_2=0 \\ m_2 \equiv m_1 \pmod{2}}}^{\infty}\sum_{N=0}^{\infty}P_{m_2}(N)q^{N+m_2}\biggr) \ ,
\ee
where the constant $c$ has been omitted, since it would just result in an overall shift of the energy spectra. Using massless single-particle energies \eqref{single-particle-energy} and setting \eqref{boltzmann} in \eqref{zwischen} then results in the partition function \eqref{partition_function} corresponding to a system of two quasi-particle species, with both species having the momentum spectrum $\N_{\geq 1}$, i.e. a multi-particle state with energy $E_l$ may consist of exactly those combinations of an even number of quasi-particles, having momenta $p_{\alpha}^i\ (i\in \left\lbrace 1,2 \right\rbrace )$, whose single-particle energies $e(p_{\alpha}^i)$ add up to $E_l$ and where the momenta $p_{\alpha}^i\in\N_{\geq 1}$ of each two of the quasi-particles in that combination are distinct unless they belong to different species, i.e. they respect the exclusion principle. Formally, these spectra belong to two free chiral fermions with periodic boundary conditions. Note in this context the physical interpretations in \cite{Kau95,Kau00}, in which the CFT for $ c_{2,1}=-2 $ is generated from a symplectic fermion, a free two-component fermion field of spin one.

\mathversion{bold}
\subsection{The $p>2$ Relatives}
\mathversion{normal}
Besides the best understood LCFT with central charge $ c_{2,1}=-2 $, some general remarks on its $ c_{p,1} $ relatives are alluded here to conclude this section.

The restrictions $ m_{p-1}+m_{p}\equiv Q \pmod{2} $ ($ Q $ denotes the total charge of the system) in \eqref{fe-1} to \eqref{fe-3} imply that the quasi-particles $ p-1 $ and $ p $ are charged under a $ {\mathbb Z}_2 $ subgroup of the full symmetry of the $ D_p $ Dynkin diagram \cite{KKMM93a}, while all the others are neutral. This charge reflects the $\mathfrak{su}(2)$ structure carried by the triplet $\mathcal{W}$-algebra such that all representations must have ground states, which are either $\mathfrak{su}(2)$ singlets or $\mathfrak{su}(2)$ doublets. In comparison to the $ c_{2,1}=-2 $ model, there exist $ p $ quasi-particles in each member of the $ c_{p,1} $ series, exactly two of which can only be created in pairs, while the others do not have this restriction. 
These observations suggest the following conjecture: The $c_{p,1}$ theories might possess a realization in terms of free fermions such that they are generated by one pair of symplectic fermions and $p-2$ ordinary fermions. Such realizations are unknown so far, except for the well-understood case $p=2$, and are a very interesting direction of future research.

Contrary to the case of $ p=2 $, the quasi-particles do not decouple here: The minimal momenta for the quasi-particle species, which are given in \eqref{pmin}, depend on the numbers of quasi-particles of the different species in the state. But as in the case of $p=2$, the momentum spectra are not bounded from above.
\section{Conclusion}

The results of our article provide further evidence for the well-definedness of the logarithmic conformal field theories corresponding to central charge $c_{p,1}$:

Despite the inhomogeneous structure of the bosonic character expressions in terms of modular forms, there exist fermionic quasi-particle sum representations with the same matrix $A$ (cf. \eqref{nahm}) for all characters of each $c_{p,1}$ model. In particular, the matrix $A$ is the inverse of the Cartan matrix of the simply-laced Lie algebra $D_p$, where $p=2$ can be understood as the degenerate case. Therefore, those expressions fit well into the known scheme of fermionic character expressions for other conformal field theories.

Physically, in each $c_{p,1}$ model, this indicates that there is a set of $p-2$ fermionic quasi-particle species, the members of which may be combined freely - nevertheless obeying Pauli's exclusion principle - in building an arbitrary multi-particle state, and additionally a set of two species, the members of which may only appear in an even or odd number, depending on the sector of the theory. In all cases except $p=2$, the possible quasi-particle momenta obey non-trivial restrictions \eqref{pmin} for their minimum momenta, depending on the numbers of quasi-particles of each species in the state. Moreover, since the fermionic character expressions are of the form \eqref{nahm} for all $p\geq 2$, the momentum spectra are unbounded from above.
Being rational CFTs \cite{GK96b,CF06}, it is furthermore satisfying that the fermionic character expressions of the outlined theories - although they are non-minimal models on the edge of the conformal grid - lead correctly to a well-known infinite set of dilogarithm identities, which supports the fermionic expressions for the characters of the $c_{p,1}$ models that we present in this article.
\\

\textbf {Acknowledgements.}
The authors would like to thank Andreas Recknagel, Anne Schilling and Trevor Welsh for useful comments and Kirsten Vogeler for carefully reading the document.
The work of MF is partially supported by the European Union network HPRN-CT-2002-00325 (EUCLID).

\bibliographystyle{grabow_koehn}
\bibliography{/home/itp/koehn/mytex/bib/master}

\end{document}